\documentclass[11pt]{article}
\usepackage{graphicx,verbatim,array,palatino,natbib}
\usepackage{amsmath,amssymb,amsfonts,ifthen,setspace,color}
\usepackage{multirow, multicol}
\usepackage{ulem}
\newboolean{ShowFigures}
\newboolean{UnBlinded}
\newboolean{DoubleSpaced}
\setboolean{ShowFigures}{true}
\setboolean{UnBlinded}{true}
\setboolean{DoubleSpaced}{false}
\usepackage{amsmath,amssymb,amsfonts}

\def\thick#1{\hbox{\rlap{$#1$}\kern0.25pt\rlap{$#1$}\kern0.25pt$#1$}}

\def\thick#1{\hbox{\rlap{$#1$}\kern0.25pt\rlap{$#1$}\kern0.25pt$#1$}}
\def\smbalpha{{\thick{\scriptstyle{\alpha}}}}

\def\smbalpha{\widehat{\smbalpha}}

\def\hbar{{\overline h}}

\def\simiid{\stackrel{{\tiny \mbox{i.i.d.}}}{\sim}}

\def\beq{\begin{equation}}
\def\eeq{\end{equation}}
\def\bev{\begin{verbatim}}
\def\eev{\end{verbatim}}

\def\jump{\vskip3mm\noindent}
\def\lboxit#1{\vbox{\hrule\hbox{\vrule\kern6pt
      \vbox{\kern6pt#1\kern6pt}\kern6pt\vrule}\hrule}}

\def\thickboxit#1{\vbox{{\hrule height 1mm}\hbox{{\vrule width 1mm}\kern6pt
          \vbox{\kern6pt#1\kern6pt}\kern6pt{\vrule width 1mm}}
               {\hrule height 1mm}}}
\def\colthickboxit#1{\vbox{{\textcolor{brown}{\hrule height 3mm}}
          \hbox{{\textcolor{brown}{\vrule width 3mm}}\kern6pt
          \vbox{\kern6pt#1\kern6pt}\kern6pt{\textcolor{brown}{\vrule width 3mm}}}
               {\hrule height 5mm}}}

\def\beq{\begin{eqnarray}}
\def\eeq{\end{eqnarray}}
\def\beqn{\begin{eqnarray*}}
\def\eeqn{\end{eqnarray*}}

\def\bse{\begin{eqnarray*}}
\def\ese{\end{eqnarray*}}
\def\raybe{\begin{eqnarray}}
\def\rayee{\end{eqnarray}}

\def\fat#1{\hbox{\rlap{$#1$}\kern0.25pt\rlap{$#1$}\kern0.25pt$#1$}}

\def\bse{\begin{eqnarray*}}
\def\ese{\end{eqnarray*}}
\def\raybe{\begin{eqnarray}}
\def\rayee{\end{eqnarray}}

\def\pe+{p_{{\scriptscriptstyle{\rm E}+}}}
\def\pg+{p_{{\scriptscriptstyle{\rm G}+}}}

\def\exerDBtags#1#2{\null}
{\begin{list}
{\bulletcolour$\bullet$}
{\setlength{\leftmargin}{30mm}\setlength{\itemsep}{0.5ex}}\sf}
{\end{list}\normalsize}
{\begin{list}{\small\bulletcolour$\bullet$}
{\setlength{\leftmargin}{0.75in}\setlength{\itemsep}{0.5ex}}\small}
{\end{list}\normalsize}
\definecolor{LightGreen}{rgb}{0.5657059,0.9333333,0.5647059}
\definecolor{gold}{rgb}{1.0,0.84,0.0}
\definecolor{LemonChiffon}{rgb}{1.0,0.98,0.9}
\definecolor{pink}{rgb}{1,0.75,0.8}
\definecolor{DeepPink}{rgb}{1.0,0.078,0.576}
\definecolor{ltyellow}{cmyk}{0,0,.4,0}
\definecolor{orange}{rgb}{1.0,0.647,0.0}
\definecolor{LightCyan}{rgb}{0.8,1.0,1.0}
\definecolor{lilac}{rgb}{1.0,0.9,1.0}
\definecolor{brown}{rgb}{0.35,0.2,0.05}
\definecolor{LightBlue}{rgb}{0.8,0.85,1}
\definecolor{grey}{gray}{0.50}
\definecolor{DarkGreen}{rgb}{0.2,0.7,0.2}
\definecolor{PaleGreen}{rgb}{0.88,1,0.88}
\definecolor{Salmon}{cmyk}{0,0.53,0.38,0}
\definecolor{ruppertgreen}{rgb}{0,.7,.25}
\definecolor{snow}{rgb}{1,0.98,0.98}
\definecolor{GhostWhite}{rgb}{0.973,0.973,1}
\definecolor{WhiteSmoke}{rgb}{0.961,0.961,0.961}
\definecolor{gainsboro}{rgb}{0.863,0.863,0.863}
\definecolor{FloralWhite}{rgb}{1,0.98,0.941}
\definecolor{OldLace}{rgb}{0.992,0.961,0.902}
\definecolor{linen}{rgb}{0.98,0.941,0.902}
\definecolor{AntiqueWhite}{rgb}{0.98,0.922,0.843}
\definecolor{PapayaWhip}{rgb}{1,0.937,0.835}
\definecolor{BlanchedAlmond}{rgb}{1,0.922,0.804}
\definecolor{bisque}{rgb}{1,0.894,0.769}
\definecolor{PeachPuff}{rgb}{1,0.855,0.725}
\definecolor{NavajoWhite}{rgb}{1,0.871,0.678}
\definecolor{moccasin}{rgb}{1,0.894,0.71}
\definecolor{cornsilk}{rgb}{1,0.973,0.863}
\definecolor{ivory}{rgb}{1,1,0.941}
\definecolor{seashell}{rgb}{1,0.961,0.933}
\definecolor{honeydew}{rgb}{0.941,1,0.941}
\definecolor{MintCream}{rgb}{0.961,1,0.98}
\definecolor{azure}{rgb}{0.941,1,1}
\definecolor{AliceBlue}{rgb}{0.941,0.973,1}
\definecolor{lavender}{rgb}{0.902,0.902,0.98}
\definecolor{LavenderBlush}{rgb}{1,0.941,0.961}
\definecolor{MistyRose}{rgb}{1,0.894,0.882}
\definecolor{white}{rgb}{1,1,1}
\definecolor{black}{rgb}{0,0,0}
\definecolor{DarkSlateGrey}{rgb}{0.184,0.31,0.31}
\definecolor{DimGrey}{rgb}{0.412,0.412,0.412}
\definecolor{SlateGrey}{rgb}{0.439,0.502,0.565}
\definecolor{LightSlateGrey}{rgb}{0.467,0.533,0.6}
\definecolor{grey}{rgb}{0.745,0.745,0.745}
\definecolor{LightGrey}{rgb}{0.827,0.827,0.827}
\definecolor{MidnightBlue}{rgb}{0.098,0.098,0.439}
\definecolor{navy}{rgb}{0,0,0.502}
\definecolor{NavyBlue}{rgb}{0,0,0.502}
\definecolor{CornflowerBlue}{rgb}{0.392,0.584,0.929}
\definecolor{DarkSlateBlue}{rgb}{0.282,0.239,0.545}
\definecolor{SlateBlue}{rgb}{0.416,0.353,0.804}
\definecolor{MediumSlateBlue}{rgb}{0.482,0.408,0.933}
\definecolor{LightSlateBlue}{rgb}{0.518,0.439,1}
\definecolor{MediumBlue}{rgb}{0,0,0.804}
\definecolor{RoyalBlue}{rgb}{0.255,0.412,0.882}
\definecolor{blue}{rgb}{0,0,1}
\definecolor{DodgerBlue}{rgb}{0.118,0.565,1}
\definecolor{DeepSkyBlue}{rgb}{0,0.749,1}
\definecolor{SkyBlue}{rgb}{0.529,0.808,0.922}
\definecolor{LightSkyBlue}{rgb}{0.529,0.808,0.98}
\definecolor{SteelBlue}{rgb}{0.275,0.51,0.706}
\definecolor{LightSteelBlue}{rgb}{0.69,0.769,0.871}
\definecolor{LightBlue}{rgb}{0.678,0.847,0.902}
\definecolor{PowderBlue}{rgb}{0.69,0.878,0.902}
\definecolor{PaleTurquoise}{rgb}{0.686,0.933,0.933}
\definecolor{DarkTurquoise}{rgb}{0,0.808,0.82}
\definecolor{MediumTurquoise}{rgb}{0.282,0.82,0.8}
\definecolor{turquoise}{rgb}{0.251,0.878,0.816}
\definecolor{cyan}{rgb}{0,1,1}
\definecolor{LightCyan}{rgb}{0.878,1,1}
\definecolor{CadetBlue}{rgb}{0.373,0.62,0.627}
\definecolor{MediumAquamarine}{rgb}{0.4,0.804,0.667}
\definecolor{aquamarine}{rgb}{0.498,1,0.831}
\definecolor{DarkGreen}{rgb}{0,0.392,0}
\definecolor{DarkOliveGreen}{rgb}{0.333,0.42,0.184}
\definecolor{DarkSeaGreen}{rgb}{0.561,0.737,0.561}
\definecolor{SeaGreen}{rgb}{0.18,0.545,0.341}
\definecolor{MediumSeaGreen}{rgb}{0.235,0.702,0.443}
\definecolor{LightSeaGreen}{rgb}{0.125,0.698,0.667}
\definecolor{PaleGreen}{rgb}{0.596,0.984,0.596}
\definecolor{SpringGreen}{rgb}{0,1,0.498}
\definecolor{LawnGreen}{rgb}{0.486,0.988,0}
\definecolor{green}{rgb}{0,1,0}
\definecolor{chartreuse}{rgb}{0.498,1,0}
\definecolor{MediumSpringGreen}{rgb}{0,0.98,0.604}
\definecolor{GreenYellow}{rgb}{0.678,1,0.184}
\definecolor{LimeGreen}{rgb}{0.196,0.804,0.196}
\definecolor{YellowGreen}{rgb}{0.604,0.804,0.196}
\definecolor{ForestGreen}{rgb}{0.133,0.545,0.133}
\definecolor{OliveDrab}{rgb}{0.42,0.557,0.137}
\definecolor{DarkKhaki}{rgb}{0.741,0.718,0.42}
\definecolor{khaki}{rgb}{0.941,0.902,0.549}
\definecolor{PaleGoldenrod}{rgb}{0.933,0.91,0.667}
\definecolor{LightGoldenrodYellow}{rgb}{0.98,0.98,0.824}
\definecolor{LightYellow}{rgb}{1,1,0.878}
\definecolor{yellow}{rgb}{1,1,0}
\definecolor{gold}{rgb}{1,0.843,0}
\definecolor{LightGoldenrod}{rgb}{0.933,0.867,0.51}
\definecolor{goldenrod}{rgb}{0.855,0.647,0.125}
\definecolor{DarkGoldenrod}{rgb}{0.722,0.525,0.043}
\definecolor{RosyBrown}{rgb}{0.737,0.561,0.561}
\definecolor{IndianRed}{rgb}{0.804,0.361,0.361}
\definecolor{SaddleBrown}{rgb}{0.545,0.271,0.075}
\definecolor{sienna}{rgb}{0.627,0.322,0.176}
\definecolor{peru}{rgb}{0.804,0.522,0.247}
\definecolor{burlywood}{rgb}{0.871,0.722,0.529}
\definecolor{beige}{rgb}{0.961,0.961,0.863}
\definecolor{wheat}{rgb}{0.961,0.871,0.702}
\definecolor{SandyBrown}{rgb}{0.957,0.643,0.376}
\definecolor{tan}{rgb}{0.824,0.706,0.549}
\definecolor{chocolate}{rgb}{0.824,0.412,0.118}
\definecolor{firebrick}{rgb}{0.698,0.133,0.133}
\definecolor{brown}{rgb}{0.647,0.165,0.165}
\definecolor{DarkSalmon}{rgb}{0.914,0.588,0.478}
\definecolor{salmon}{rgb}{0.98,0.502,0.447}
\definecolor{LightSalmon}{rgb}{1,0.627,0.478}
\definecolor{orange}{rgb}{1,0.647,0}
\definecolor{DarkOrange}{rgb}{1,0.549,0}
\definecolor{coral}{rgb}{1,0.498,0.314}
\definecolor{LightCoral}{rgb}{0.941,0.502,0.502}
\definecolor{tomato}{rgb}{1,0.388,0.278}
\definecolor{OrangeRed}{rgb}{1,0.271,0}
\definecolor{red}{rgb}{1,0,0}
\definecolor{HotPink}{rgb}{1,0.412,0.706}
\definecolor{DeepPink}{rgb}{1,0.078,0.576}
\definecolor{pink}{rgb}{1,0.753,0.796}
\definecolor{LightPink}{rgb}{1,0.714,0.757}
\definecolor{PaleVioletRed}{rgb}{0.859,0.439,0.576}
\definecolor{maroon}{rgb}{0.69,0.188,0.376}
\definecolor{MediumVioletRed}{rgb}{0.78,0.082,0.522}
\definecolor{VioletRed}{rgb}{0.816,0.125,0.565}
\definecolor{magenta}{rgb}{1,0,1}
\definecolor{violet}{rgb}{0.933,0.51,0.933}
\definecolor{plum}{rgb}{0.867,0.627,0.867}
\definecolor{orchid}{rgb}{0.855,0.439,0.839}
\definecolor{MediumOrchid}{rgb}{0.729,0.333,0.827}
\definecolor{DarkOrchid}{rgb}{0.6,0.196,0.8}
\definecolor{DarkViolet}{rgb}{0.58,0,0.827}
\definecolor{BlueViolet}{rgb}{0.541,0.169,0.886}
\definecolor{purple}{rgb}{0.627,0.125,0.941}
\definecolor{MediumPurple}{rgb}{0.576,0.439,0.859}
\definecolor{thistle}{rgb}{0.847,0.749,0.847}
\definecolor{snow1}{rgb}{1,0.98,0.98}
\definecolor{snow2}{rgb}{0.933,0.914,0.914}
\definecolor{snow3}{rgb}{0.804,0.788,0.788}
\definecolor{snow4}{rgb}{0.545,0.537,0.537}
\definecolor{seashell1}{rgb}{1,0.961,0.933}
\definecolor{seashell2}{rgb}{0.933,0.898,0.871}
\definecolor{seashell3}{rgb}{0.804,0.773,0.749}
\definecolor{seashell4}{rgb}{0.545,0.525,0.51}
\definecolor{AntiqueWhite1}{rgb}{1,0.937,0.859}
\definecolor{AntiqueWhite2}{rgb}{0.933,0.875,0.8}
\definecolor{AntiqueWhite3}{rgb}{0.804,0.753,0.69}
\definecolor{AntiqueWhite4}{rgb}{0.545,0.514,0.471}
\definecolor{bisque1}{rgb}{1,0.894,0.769}
\definecolor{bisque2}{rgb}{0.933,0.835,0.718}
\definecolor{bisque3}{rgb}{0.804,0.718,0.62}
\definecolor{bisque4}{rgb}{0.545,0.49,0.42}
\definecolor{PeachPuff1}{rgb}{1,0.855,0.725}
\definecolor{PeachPuff2}{rgb}{0.933,0.796,0.678}
\definecolor{PeachPuff3}{rgb}{0.804,0.686,0.584}
\definecolor{PeachPuff4}{rgb}{0.545,0.467,0.396}
\definecolor{NavajoWhite1}{rgb}{1,0.871,0.678}
\definecolor{NavajoWhite2}{rgb}{0.933,0.812,0.631}
\definecolor{NavajoWhite3}{rgb}{0.804,0.702,0.545}
\definecolor{NavajoWhite4}{rgb}{0.545,0.475,0.369}
\definecolor{LemonChiffon1}{rgb}{1,0.98,0.804}
\definecolor{LemonChiffon2}{rgb}{0.933,0.914,0.749}
\definecolor{LemonChiffon3}{rgb}{0.804,0.788,0.647}
\definecolor{LemonChiffon4}{rgb}{0.545,0.537,0.439}
\definecolor{cornsilk1}{rgb}{1,0.973,0.863}
\definecolor{cornsilk2}{rgb}{0.933,0.91,0.804}
\definecolor{cornsilk3}{rgb}{0.804,0.784,0.694}
\definecolor{cornsilk4}{rgb}{0.545,0.533,0.471}
\definecolor{ivory1}{rgb}{1,1,0.941}
\definecolor{ivory2}{rgb}{0.933,0.933,0.878}
\definecolor{ivory3}{rgb}{0.804,0.804,0.757}
\definecolor{ivory4}{rgb}{0.545,0.545,0.514}
\definecolor{honeydew1}{rgb}{0.941,1,0.941}
\definecolor{honeydew2}{rgb}{0.878,0.933,0.878}
\definecolor{honeydew3}{rgb}{0.757,0.804,0.757}
\definecolor{honeydew4}{rgb}{0.514,0.545,0.514}
\definecolor{LavenderBlush1}{rgb}{1,0.941,0.961}
\definecolor{LavenderBlush2}{rgb}{0.933,0.878,0.898}
\definecolor{LavenderBlush3}{rgb}{0.804,0.757,0.773}
\definecolor{LavenderBlush4}{rgb}{0.545,0.514,0.525}
\definecolor{MistyRose2}{rgb}{0.933,0.835,0.824}
\definecolor{MistyRose3}{rgb}{0.804,0.718,0.71}
\definecolor{MistyRose4}{rgb}{0.545,0.49,0.482}
\definecolor{azure1}{rgb}{0.941,1,1}
\definecolor{azure2}{rgb}{0.878,0.933,0.933}
\definecolor{azure3}{rgb}{0.757,0.804,0.804}
\definecolor{azure4}{rgb}{0.514,0.545,0.545}
\definecolor{SlateBlue1}{rgb}{0.514,0.435,1}
\definecolor{SlateBlue2}{rgb}{0.478,0.404,0.933}
\definecolor{SlateBlue3}{rgb}{0.412,0.349,0.804}
\definecolor{SlateBlue4}{rgb}{0.278,0.235,0.545}
\definecolor{RoyalBlue1}{rgb}{0.282,0.463,1}
\definecolor{RoyalBlue2}{rgb}{0.263,0.431,0.933}
\definecolor{RoyalBlue3}{rgb}{0.227,0.373,0.804}
\definecolor{RoyalBlue4}{rgb}{0.153,0.251,0.545}
\definecolor{blue1}{rgb}{0,0,1}
\definecolor{blue2}{rgb}{0,0,0.933}
\definecolor{blue3}{rgb}{0,0,0.804}
\definecolor{blue4}{rgb}{0,0,0.545}
\definecolor{DodgerBlue1}{rgb}{0.118,0.565,1}
\definecolor{DodgerBlue2}{rgb}{0.11,0.525,0.933}
\definecolor{DodgerBlue3}{rgb}{0.094,0.455,0.804}
\definecolor{DodgerBlue4}{rgb}{0.063,0.306,0.545}
\definecolor{SteelBlue1}{rgb}{0.388,0.722,1}
\definecolor{SteelBlue2}{rgb}{0.361,0.675,0.933}
\definecolor{SteelBlue3}{rgb}{0.31,0.58,0.804}
\definecolor{SteelBlue4}{rgb}{0.212,0.392,0.545}
\definecolor{DeepSkyBlue1}{rgb}{0,0.749,1}
\definecolor{DeepSkyBlue2}{rgb}{0,0.698,0.933}
\definecolor{DeepSkyBlue3}{rgb}{0,0.604,0.804}
\definecolor{DeepSkyBlue4}{rgb}{0,0.408,0.545}
\definecolor{SkyBlue1}{rgb}{0.529,0.808,1}
\definecolor{SkyBlue2}{rgb}{0.494,0.753,0.933}
\definecolor{SkyBlue3}{rgb}{0.424,0.651,0.804}
\definecolor{SkyBlue4}{rgb}{0.29,0.439,0.545}
\definecolor{LightSkyBlue1}{rgb}{0.69,0.886,1}
\definecolor{LightSkyBlue2}{rgb}{0.643,0.827,0.933}
\definecolor{LightSkyBlue3}{rgb}{0.553,0.714,0.804}
\definecolor{LightSkyBlue4}{rgb}{0.376,0.482,0.545}
\definecolor{SlateGray1}{rgb}{0.776,0.886,1}
\definecolor{SlateGray2}{rgb}{0.725,0.827,0.933}
\definecolor{SlateGray3}{rgb}{0.624,0.714,0.804}
\definecolor{SlateGray4}{rgb}{0.424,0.482,0.545}
\definecolor{LightSteelBlue1}{rgb}{0.792,0.882,1}
\definecolor{LightSteelBlue2}{rgb}{0.737,0.824,0.933}
\definecolor{LightSteelBlue3}{rgb}{0.635,0.71,0.804}
\definecolor{LightSteelBlue4}{rgb}{0.431,0.482,0.545}
\definecolor{LightBlue1}{rgb}{0.749,0.937,1}
\definecolor{LightBlue2}{rgb}{0.698,0.875,0.933}
\definecolor{LightBlue3}{rgb}{0.604,0.753,0.804}
\definecolor{LightBlue4}{rgb}{0.408,0.514,0.545}
\definecolor{LightCyan1}{rgb}{0.878,1,1}
\definecolor{LightCyan2}{rgb}{0.82,0.933,0.933}
\definecolor{LightCyan3}{rgb}{0.706,0.804,0.804}
\definecolor{LightCyan4}{rgb}{0.478,0.545,0.545}
\definecolor{PaleTurquoise1}{rgb}{0.733,1,1}
\definecolor{PaleTurquoise2}{rgb}{0.682,0.933,0.933}
\definecolor{PaleTurquoise3}{rgb}{0.588,0.804,0.804}
\definecolor{PaleTurquoise4}{rgb}{0.4,0.545,0.545}
\definecolor{CadetBlue1}{rgb}{0.596,0.961,1}
\definecolor{CadetBlue2}{rgb}{0.557,0.898,0.933}
\definecolor{CadetBlue3}{rgb}{0.478,0.773,0.804}
\definecolor{CadetBlue4}{rgb}{0.325,0.525,0.545}
\definecolor{turquoise1}{rgb}{0,0.961,1}
\definecolor{turquoise2}{rgb}{0,0.898,0.933}
\definecolor{turquoise3}{rgb}{0,0.773,0.804}
\definecolor{turquoise4}{rgb}{0,0.525,0.545}
\definecolor{cyan1}{rgb}{0,1,1}
\definecolor{cyan2}{rgb}{0,0.933,0.933}
\definecolor{cyan3}{rgb}{0,0.804,0.804}
\definecolor{cyan4}{rgb}{0,0.545,0.545}
\definecolor{DarkSlateGray1}{rgb}{0.592,1,1}
\definecolor{DarkSlateGray2}{rgb}{0.553,0.933,0.933}
\definecolor{DarkSlateGray3}{rgb}{0.475,0.804,0.804}
\definecolor{DarkSlateGray4}{rgb}{0.322,0.545,0.545}
\definecolor{aquamarine1}{rgb}{0.498,1,0.831}
\definecolor{aquamarine2}{rgb}{0.463,0.933,0.776}
\definecolor{aquamarine3}{rgb}{0.4,0.804,0.667}
\definecolor{aquamarine4}{rgb}{0.271,0.545,0.455}
\definecolor{DarkSeaGreen1}{rgb}{0.757,1,0.757}
\definecolor{DarkSeaGreen2}{rgb}{0.706,0.933,0.706}
\definecolor{DarkSeaGreen3}{rgb}{0.608,0.804,0.608}
\definecolor{DarkSeaGreen4}{rgb}{0.412,0.545,0.412}
\definecolor{SeaGreen1}{rgb}{0.329,1,0.624}
\definecolor{SeaGreen2}{rgb}{0.306,0.933,0.58}
\definecolor{SeaGreen3}{rgb}{0.263,0.804,0.502}
\definecolor{SeaGreen4}{rgb}{0.18,0.545,0.341}
\definecolor{PaleGreen1}{rgb}{0.604,1,0.604}
\definecolor{PaleGreen2}{rgb}{0.565,0.933,0.565}
\definecolor{PaleGreen3}{rgb}{0.486,0.804,0.486}
\definecolor{PaleGreen4}{rgb}{0.329,0.545,0.329}
\definecolor{SpringGreen1}{rgb}{0,1,0.498}
\definecolor{SpringGreen2}{rgb}{0,0.933,0.463}
\definecolor{SpringGreen3}{rgb}{0,0.804,0.4}
\definecolor{SpringGreen4}{rgb}{0,0.545,0.271}
\definecolor{green1}{rgb}{0,1,0}
\definecolor{green2}{rgb}{0,0.933,0}
\definecolor{green3}{rgb}{0,0.804,0}
\definecolor{green4}{rgb}{0,0.545,0}
\definecolor{chartreuse1}{rgb}{0.498,1,0}
\definecolor{chartreuse2}{rgb}{0.463,0.933,0}
\definecolor{chartreuse3}{rgb}{0.4,0.804,0}
\definecolor{chartreuse4}{rgb}{0.271,0.545,0}
\definecolor{OliveDrab1}{rgb}{0.753,1,0.243}
\definecolor{OliveDrab2}{rgb}{0.702,0.933,0.227}
\definecolor{OliveDrab3}{rgb}{0.604,0.804,0.196}
\definecolor{OliveDrab4}{rgb}{0.412,0.545,0.133}
\definecolor{DarkOliveGreen1}{rgb}{0.792,1,0.439}
\definecolor{DarkOliveGreen2}{rgb}{0.737,0.933,0.408}
\definecolor{DarkOliveGreen3}{rgb}{0.635,0.804,0.353}
\definecolor{DarkOliveGreen4}{rgb}{0.431,0.545,0.239}
\definecolor{khaki1}{rgb}{1,0.965,0.561}
\definecolor{khaki2}{rgb}{0.933,0.902,0.522}
\definecolor{khaki3}{rgb}{0.804,0.776,0.451}
\definecolor{khaki4}{rgb}{0.545,0.525,0.306}
\definecolor{LightGoldenrod1}{rgb}{1,0.925,0.545}
\definecolor{LightGoldenrod2}{rgb}{0.933,0.863,0.51}
\definecolor{LightGoldenrod3}{rgb}{0.804,0.745,0.439}
\definecolor{LightGoldenrod4}{rgb}{0.545,0.506,0.298}
\definecolor{LightYellow1}{rgb}{1,1,0.878}
\definecolor{LightYellow2}{rgb}{0.933,0.933,0.82}
\definecolor{LightYellow3}{rgb}{0.804,0.804,0.706}
\definecolor{LightYellow4}{rgb}{0.545,0.545,0.478}
\definecolor{yellow1}{rgb}{1,1,0}
\definecolor{yellow2}{rgb}{0.933,0.933,0}
\definecolor{yellow3}{rgb}{0.804,0.804,0}
\definecolor{yellow4}{rgb}{0.545,0.545,0}
\definecolor{gold1}{rgb}{1,0.843,0}
\definecolor{gold2}{rgb}{0.933,0.788,0}
\definecolor{gold3}{rgb}{0.804,0.678,0}
\definecolor{gold4}{rgb}{0.545,0.459,0}
\definecolor{goldenrod1}{rgb}{1,0.757,0.145}
\definecolor{goldenrod2}{rgb}{0.933,0.706,0.133}
\definecolor{goldenrod3}{rgb}{0.804,0.608,0.114}
\definecolor{goldenrod4}{rgb}{0.545,0.412,0.078}
\definecolor{DarkGoldenrod1}{rgb}{1,0.725,0.059}
\definecolor{DarkGoldenrod2}{rgb}{0.933,0.678,0.055}
\definecolor{DarkGoldenrod3}{rgb}{0.804,0.584,0.047}
\definecolor{DarkGoldenrod4}{rgb}{0.545,0.396,0.031}
\definecolor{RosyBrown1}{rgb}{1,0.757,0.757}
\definecolor{RosyBrown2}{rgb}{0.933,0.706,0.706}
\definecolor{RosyBrown3}{rgb}{0.804,0.608,0.608}
\definecolor{RosyBrown4}{rgb}{0.545,0.412,0.412}
\definecolor{IndianRed1}{rgb}{1,0.416,0.416}
\definecolor{IndianRed2}{rgb}{0.933,0.388,0.388}
\definecolor{IndianRed3}{rgb}{0.804,0.333,0.333}
\definecolor{IndianRed4}{rgb}{0.545,0.227,0.227}
\definecolor{sienna1}{rgb}{1,0.51,0.278}
\definecolor{sienna2}{rgb}{0.933,0.475,0.259}
\definecolor{sienna3}{rgb}{0.804,0.408,0.224}
\definecolor{sienna4}{rgb}{0.545,0.278,0.149}
\definecolor{burlywood1}{rgb}{1,0.827,0.608}
\definecolor{burlywood2}{rgb}{0.933,0.773,0.569}
\definecolor{burlywood3}{rgb}{0.804,0.667,0.49}
\definecolor{burlywood4}{rgb}{0.545,0.451,0.333}
\definecolor{wheat1}{rgb}{1,0.906,0.729}
\definecolor{wheat2}{rgb}{0.933,0.847,0.682}
\definecolor{wheat3}{rgb}{0.804,0.729,0.588}
\definecolor{wheat4}{rgb}{0.545,0.494,0.4}
\definecolor{tan1}{rgb}{1,0.647,0.31}
\definecolor{tan2}{rgb}{0.933,0.604,0.286}
\definecolor{tan3}{rgb}{0.804,0.522,0.247}
\definecolor{tan4}{rgb}{0.545,0.353,0.169}
\definecolor{chocolate1}{rgb}{1,0.498,0.141}
\definecolor{chocolate2}{rgb}{0.933,0.463,0.129}
\definecolor{chocolate3}{rgb}{0.804,0.4,0.114}
\definecolor{chocolate4}{rgb}{0.545,0.271,0.075}
\definecolor{firebrick1}{rgb}{1,0.188,0.188}
\definecolor{firebrick2}{rgb}{0.933,0.173,0.173}
\definecolor{firebrick3}{rgb}{0.804,0.149,0.149}
\definecolor{firebrick4}{rgb}{0.545,0.102,0.102}
\definecolor{brown1}{rgb}{1,0.251,0.251}
\definecolor{brown2}{rgb}{0.933,0.231,0.231}
\definecolor{brown3}{rgb}{0.804,0.2,0.2}
\definecolor{brown4}{rgb}{0.545,0.137,0.137}
\definecolor{salmon1}{rgb}{1,0.549,0.412}
\definecolor{salmon2}{rgb}{0.933,0.51,0.384}
\definecolor{salmon3}{rgb}{0.804,0.439,0.329}
\definecolor{salmon4}{rgb}{0.545,0.298,0.224}
\definecolor{LightSalmon1}{rgb}{1,0.627,0.478}
\definecolor{LightSalmon2}{rgb}{0.933,0.584,0.447}
\definecolor{LightSalmon3}{rgb}{0.804,0.506,0.384}
\definecolor{LightSalmon4}{rgb}{0.545,0.341,0.259}
\definecolor{orange1}{rgb}{1,0.647,0}
\definecolor{orange2}{rgb}{0.933,0.604,0}
\definecolor{orange3}{rgb}{0.804,0.522,0}
\definecolor{orange4}{rgb}{0.545,0.353,0}
\definecolor{DarkOrange1}{rgb}{1,0.498,0}
\definecolor{DarkOrange2}{rgb}{0.933,0.463,0}
\definecolor{DarkOrange3}{rgb}{0.804,0.4,0}
\definecolor{DarkOrange4}{rgb}{0.545,0.271,0}
\definecolor{coral1}{rgb}{1,0.447,0.337}
\definecolor{coral2}{rgb}{0.933,0.416,0.314}
\definecolor{coral3}{rgb}{0.804,0.357,0.271}
\definecolor{coral4}{rgb}{0.545,0.243,0.184}
\definecolor{tomato1}{rgb}{1,0.388,0.278}
\definecolor{tomato2}{rgb}{0.933,0.361,0.259}
\definecolor{tomato3}{rgb}{0.804,0.31,0.224}
\definecolor{tomato4}{rgb}{0.545,0.212,0.149}
\definecolor{OrangeRed1}{rgb}{1,0.271,0}
\definecolor{OrangeRed2}{rgb}{0.933,0.251,0}
\definecolor{OrangeRed3}{rgb}{0.804,0.216,0}
\definecolor{OrangeRed4}{rgb}{0.545,0.145,0}
\definecolor{red1}{rgb}{1,0,0}
\definecolor{red2}{rgb}{0.933,0,0}
\definecolor{red3}{rgb}{0.804,0,0}
\definecolor{red4}{rgb}{0.545,0,0}
\definecolor{DeepPink1}{rgb}{1,0.078,0.576}
\definecolor{DeepPink2}{rgb}{0.933,0.071,0.537}
\definecolor{DeepPink3}{rgb}{0.804,0.063,0.463}
\definecolor{DeepPink4}{rgb}{0.545,0.039,0.314}
\definecolor{HotPink1}{rgb}{1,0.431,0.706}
\definecolor{HotPink2}{rgb}{0.933,0.416,0.655}
\definecolor{HotPink3}{rgb}{0.804,0.376,0.565}
\definecolor{HotPink4}{rgb}{0.545,0.227,0.384}
\definecolor{pink1}{rgb}{1,0.71,0.773}
\definecolor{pink2}{rgb}{0.933,0.663,0.722}
\definecolor{pink3}{rgb}{0.804,0.569,0.62}
\definecolor{pink4}{rgb}{0.545,0.388,0.424}
\definecolor{LightPink1}{rgb}{1,0.682,0.725}
\definecolor{LightPink2}{rgb}{0.933,0.635,0.678}
\definecolor{LightPink3}{rgb}{0.804,0.549,0.584}
\definecolor{LightPink4}{rgb}{0.545,0.373,0.396}
\definecolor{PaleVioletRed1}{rgb}{1,0.51,0.671}
\definecolor{PaleVioletRed2}{rgb}{0.933,0.475,0.624}
\definecolor{PaleVioletRed3}{rgb}{0.804,0.408,0.537}
\definecolor{PaleVioletRed4}{rgb}{0.545,0.278,0.365}
\definecolor{maroon1}{rgb}{1,0.204,0.702}
\definecolor{maroon2}{rgb}{0.933,0.188,0.655}
\definecolor{maroon3}{rgb}{0.804,0.161,0.565}
\definecolor{maroon4}{rgb}{0.545,0.11,0.384}
\definecolor{VioletRed1}{rgb}{1,0.243,0.588}
\definecolor{VioletRed2}{rgb}{0.933,0.227,0.549}
\definecolor{VioletRed3}{rgb}{0.804,0.196,0.471}
\definecolor{VioletRed4}{rgb}{0.545,0.133,0.322}
\definecolor{magenta1}{rgb}{1,0,1}
\definecolor{magenta2}{rgb}{0.933,0,0.933}
\definecolor{magenta3}{rgb}{0.804,0,0.804}
\definecolor{magenta4}{rgb}{0.545,0,0.545}
\definecolor{orchid1}{rgb}{1,0.514,0.98}
\definecolor{orchid2}{rgb}{0.933,0.478,0.914}
\definecolor{orchid3}{rgb}{0.804,0.412,0.788}
\definecolor{orchid4}{rgb}{0.545,0.278,0.537}
\definecolor{plum1}{rgb}{1,0.733,1}
\definecolor{plum2}{rgb}{0.933,0.682,0.933}
\definecolor{plum3}{rgb}{0.804,0.588,0.804}
\definecolor{plum4}{rgb}{0.545,0.4,0.545}
\definecolor{MediumOrchid1}{rgb}{0.878,0.4,1}
\definecolor{MediumOrchid2}{rgb}{0.82,0.373,0.933}
\definecolor{MediumOrchid3}{rgb}{0.706,0.322,0.804}
\definecolor{MediumOrchid4}{rgb}{0.478,0.216,0.545}
\definecolor{DarkOrchid1}{rgb}{0.749,0.243,1}
\definecolor{DarkOrchid2}{rgb}{0.698,0.227,0.933}
\definecolor{DarkOrchid3}{rgb}{0.604,0.196,0.804}
\definecolor{DarkOrchid4}{rgb}{0.408,0.133,0.545}
\definecolor{purple1}{rgb}{0.608,0.188,1}
\definecolor{purple2}{rgb}{0.569,0.173,0.933}
\definecolor{purple3}{rgb}{0.49,0.149,0.804}
\definecolor{purple4}{rgb}{0.333,0.102,0.545}
\definecolor{MediumPurple1}{rgb}{0.671,0.51,1}
\definecolor{MediumPurple2}{rgb}{0.624,0.475,0.933}
\definecolor{MediumPurple3}{rgb}{0.537,0.408,0.804}
\definecolor{MediumPurple4}{rgb}{0.365,0.278,0.545}
\definecolor{thistle1}{rgb}{1,0.882,1}
\definecolor{thistle2}{rgb}{0.933,0.824,0.933}
\definecolor{thistle3}{rgb}{0.804,0.71,0.804}
\definecolor{thistle4}{rgb}{0.545,0.482,0.545}
\definecolor{gray0}{rgb}{0,0,0}

\newtheorem{Proposition}{Proposition}
\newtheorem{Corollary}{Corollary}

\usepackage{hyperref}
\hypersetup{colorlinks,
linkcolor=blue,
citecolor=blue}
\begin{document}
\setcounter{page}{1}

\ifthenelse{\boolean{DoubleSpaced}}
{\setstretch{1.5}}{}

\begin{center}
{\Large\bf Orthogonality of the Mean and Error Distribution} 
\jump
{\Large\bf  in Generalized Linear Models}\footnote{Accepted for publication {\it in Communications in Statistics - Theory and Methods}, on 20 Sep 2013. DOI: 10.1080/03610926.2013.851241}

\vskip4mm
{\sc By Alan Huang}
\footnote{Corresponding Author. Address: School of Mathematics and Physics, University of Queensland, AUSTRALIA, 4072. E-mail: alan.huang@uq.edu.au. }
and {\sc Paul J. Rathouz}
\footnote{Address: K6/446a, Clinical Science Center, 600 Highland Ave, Madison, WI, USA, 53792. Phone: 608-263-1706. Email: rathouz@biostat.wisc.edu.}
\\
\vskip4mm
{\it University of Technology Sydney} and {\it University of Wisconsin--Madison}
\end{center}

\vskip3mm
\centerline{4th August, 2013}
\vskip3mm
\vskip2mm
\noindent
{\bf Abstract}
We show that the mean-model parameter is always orthogonal to the error distribution in generalized linear models. Thus, the maximum likelihood estimator of the mean-model parameter will be asymptotically efficient regardless of whether the error distribution is known completely, known up to a finite vector of parameters, or left completely unspecified, in which case the likelihood is taken to be an appropriate semiparametric likelihood. Moreover, the maximum likelihood estimator of the mean-model parameter will be asymptotically independent of the maximum likelihood estimator of the error distribution.
This generalizes some well-known results for the special cases of normal, gamma and multinomial regression models, and, perhaps more interestingly, suggests that asymptotically efficient estimation and inferences can always be obtained if the error distribution is nonparametrically estimated along with the mean. 
 In contrast, estimation and inferences using 
misspecified error distributions
or 
variance functions are generally not efficient. 

\noindent
\jump

\section{Introduction}\label{sec:intro}
It is well-known that in the normal linear regression model,
$$
Y_i | X_i \sim X_i^T\beta + \epsilon_i\ , \quad \epsilon_i \simiid N(0,\sigma^2)\ ,
$$
the mean-model parameter $\beta$ is orthogonal to the error variance $\sigma^2$ \citep[e.g.][Section 3.3]{CR1987}. 
There are two important implications of this. 
First, the maximum likelihood estimator (MLE) of $\beta$ is asymptotically efficient regardless of whether $\sigma^2$ is known or estimated simultaneously from the data. Second, the MLE of $\beta$ is independent of the MLE of $\sigma^2$, which is central to deriving the usual $t$-tests for inferences on $\beta$. (Note that orthogonal parameters are only asymptotically independent in general; finite-sample independence is special to the normal distribution.) When interest lies primarily in the mean-model, the error variance is often deemed a nuisance parameter.

Similar orthogonality results hold for other generalized linear models (GLMs). Specific examples include the gamma regression model, in which the mean-model parameter is orthogonal to a nuisance shape parameter \citep[][Section 3.2]{CR1987}, and multinomial models for polytomous data, in which the mean-model parameter is orthogonal to a nuisance vector of baseline probability masses \citep{RG2009}. Note that the orthogonality property holds for any link function. 

In each of the above settings, the error distribution is characterized by a finite vector of nuisance parameters and orthogonality is established by showing that the Fisher information matrix is block-diagonal, with the blocks corresponding to the vector of mean-model parameters and the vector of nuisance parameters. It is usually straightforward to perform these calculations on a case-by-case basis, working with the specific family of distributions under consideration, but a general result for parametric GLMs can be found in \citet{JK2004}.

When we move away from specific parametric families to consider the class of all GLMs, we find that a general orthogonality property, although expected, may not be so easy to establish. This is because such a class constitutes a semiparametric model, and it is no longer feasible to compute and examine the Fisher information matrix in the presence of an infinite-dimensional parameter. 

In this note, we show that the mean-model parameter is always orthogonal to the error distribution in GLMs, even when  
the error distribution is treated as an infinite-dimensional parameter, belonging to the space of all distributions having a Laplace transform in some neighborhood of zero. This class includes, as special cases, the classical normal, Poisson, gamma and multinomial distributions, as well as many interesting and non-standard distributions, such as the generalized Poisson distribution of \citet{WF1997} for overdispersed counts and the class of all exponential dispersion models with constant dispersion \citep{Jorg1987}. We note in Section \ref{sec:2} that this class of distributions is as large as possible for GLMs, so the result here is indeed the most general possible. 

That a general orthogonality property should hold is alluded to in \citet[][Section 6.2]{JK2004}. In that paper, the notion of orthogonality between a finite-dimensional and infinite-dimensional parameter being considered is that orthogonality holds, in the usual Fisher information matrix sense, for every finite-dimensional submodel. In this note, we use a slightly stronger notion of orthogonality, namely, that the score function for the finite-dimensional parameter is orthogonal to the nuisance tangent space of the infinite-dimensional parameter. Recalling that the nuisance tangent space is the closure of all finite-dimensional submodel tangent spaces, we see that the notion of orthogonality here implies the notion considered in \citet{JK2004}.

Our proof proceeds along the following lines. We first use an exponential tilt representation of GLMs, introduced in \citet{RG2009} and expanded upon in \citet{Huang2013}, to index any GLM by just two parameters, namely, a finite-dimensional mean-model parameter $\beta$ and an infinite-dimensional error distribution parameter $F$. The orthogonality of the two parameters is then characterized by the orthogonality of the score function for $\beta$ to the nuisance tangent space for $F$. As it turns out, the nuisance tangent space for $F$ is rather difficult to work with directly, but by embedding the model into a larger class of models, we find that the required calculations become particularly simple. 

The exponential tilt representation is derived in Section \ref{sec:2} and the general orthogonality property is proven in Section \ref{sec:3}. A connection with mean-variance models is outlined in Section \ref{sec:4}. A brief discussion of the theoretical and practical implications of the our findings is given in Section \ref{sec:5}, which concludes the note.

\section{An exponential tilt representation of generalized linear models}
\label{sec:2}

Recall that a GLM \citep{MN1989} for independent data pairs $(X_1,Y_1), \ldots, (X_n, Y_n)$ is defined by two components. First, there is a conditional mean model for the responses,
\begin{equation}
E(Y_i|X_i) = \mu(X_i^T\beta) \ ,
\label{eq:meanmodel}
\end{equation}
where $\mu$ is a user-specified inverse-link function and $\beta \in \mathbb{R}^q$ is a vector of unknown regression parameters. Second, the conditional distributions $F_i$ of each response $Y_i$ given covariate $X_i$ are assumed to come from some exponential family. Assuming the distributions $F_i$ have densities $dF_i$ with respect to some dominating measure, the second component can be written in the exponential tilt form
\begin{equation}
dF_i(y) = \exp\{b(X_i; \beta, F) + \theta(X_i; \beta, F) y\} dF(y)
\end{equation}
for some reference distribution $F$, where 
\begin{equation}
\label{eq:norm}
b(X_i; \beta, F) = -\log \int \exp\{\theta(X_i; \beta, F) y\} dF(y)
\end{equation}
is a normalizing function and, in order to satisfy (\ref{eq:meanmodel}), the tilt $\theta(X_i;\beta,F)$ is implicitly defined as the solution to the mean constraint
\begin{equation}
\mu(X_i^T\beta) = \int y \exp\{b(X_i; \beta, F) +\theta(X_i;\beta,F) y\} dF(y) \ .
\label{eq:mean}
\end{equation}

It is easy to see that the exponential tilt representation (\ref{eq:meanmodel})--(\ref{eq:mean}) encompasses all classical GLMs. For example, normal, Poisson and gamma regression models can be recovered by choosing $dF$ to be a Gaussian, Poisson or gamma kernel, respectively. The main advantage of this representation is that it naturally allows for the reference distribution $F$ to be considered as an infinite-dimensional nuisance parameter, along with the finite-dimensional parameter $\beta$, in the model. It is this novel representation that allows us to conveniently 
characterize any GLM using just the two parameters $\beta$ and $F$.

As with any GLM, the reference distribution $F$ is required to have a Laplace transform in some neighborhood of the origin so that the cumulant generating function (\ref{eq:norm}) is well-defined. Thus, the parameter space for $F$ is the class of all distributions that have a Laplace transform in some neighborhood of the origin. Note that this class of distribution functions is as large as it can be for GLMs, because any distribution outside this class cannot be used to generate a valid model.

The exponential tilt representation (\ref{eq:meanmodel})--(\ref{eq:mean}) was first introduced in \citet{RG2009}. In that paper, the representation is used to derive a useful alternative parametrization of the multinomial regression model for polytomous responses. The representation is also used in \cite{Huang2013} to motivate a semiparametric extension of GLMs for arbitrary responses.

\section{The orthogonality of parameters}
\label{sec:3}
In parametric models, orthogonality of parameters can be characterized by the Fisher information matrix being block-diagonal. In semiparametric models however, orthogonality between a finite-dimensional parameter $\beta$ and an infinite-dimensional parameter $F$ cannot be characterized in this way. Rather, it is characterized through the score function for $\beta$ being orthogonal to the nuisance tangent space for $F$. Intuitively speaking, the projection of the score function for $\beta$ on to the nuisance tangent space is a measure of the loss of information about $\beta$ due to the presence of the nuisance parameter $F$ -- this is zero if and only if the score function is orthogonal to the nuisance tangent space. Note that this general notion of orthogonality reduces to the Fisher information matrix criterion when the nuisance parameter is finite-dimensional.

Now, the loglikelihood function corresponding to model (\ref{eq:meanmodel})--(\ref{eq:mean}) is $l(\beta, F|X,Y) = \log dF(Y) + b(X; \beta, F) + \theta(X;\beta, F) Y$. Thus, the
score function for $\beta$ is given by
$$S_{\beta,F}(X,Y) := \frac{\partial}{\partial \beta} l(\beta, F) = \frac{\partial}{\partial \beta} b(X;\beta, F) + \frac{\partial}{\partial \beta} \theta(X;\beta, F) Y.$$
Implicit differentiation of the defining equations for 
$b$ and $\theta$ leads to the identities
\begin{eqnarray*}
\frac{\partial }{\partial \beta} b(X;\beta,F) = - \mu (X^T\beta) \frac{\partial }{\partial \beta} \theta(X;\beta,F)  \quad \mbox{ and } \quad
\frac{\partial }{\partial \beta} \theta(X;\beta,F) = \frac{\mu'(X^T\beta) }{V(X;\beta,F)} X\ ,
\end{eqnarray*}
where
$V(X;\beta,F) = E_{\beta,F}[(Y-\mu(X^T\beta))^2|X]$
is the conditional variance of $Y$ given $X$ under parameter value $(\beta,F)$.  The score function for $\beta$ therefore reduces to
\begin{equation}
S_{\beta,F}(X,Y) = X\frac{\mu'(X^T\beta)}{V(X;\beta,F)} \left(Y-\mu(X^T\beta)\right)  \ ,
\label{eq:betascore}
\end{equation}
which is of the same weighted least-squares form as for a parametric GLM. The difference here is that the variance function $V(X;\beta,F)$ is not known because $F$ is not specified.

The orthogonality between $\beta$ and $F$ now reduces to the score function (\ref{eq:betascore}) being orthogonal to the nuisance tangent space for $F$. Although it is not hard to derive a score function for $F$ \citep[e.g.][Section 3.3]{Huang2013}, it turns out to be rather difficult to compute the nuisance tangent space explicitly. This is also noted in \citet[][Section 6.2]{JK2004}. We can work around this, however, by embedding model (\ref{eq:meanmodel})--(\ref{eq:mean}) into a more general class of ``semiparametric restricted moment models" \citep[e.g.][Section 4.5]{Tsia2006} for which the required calculations are much easier. This class is given by
\begin{equation}
\label{eq:srmm}
Y = \mu(X,\beta) + \epsilon \ ,
\end{equation}
where the conditional distribution of $\epsilon$ given $X$ is specified only up to the moment condition $E(\epsilon|X) = 0$. It is clear that the semiparametric extension (\ref{eq:meanmodel})--(\ref{eq:mean}) is a subclass of the restricted moment model, with $\mu(X,\beta) = \mu(X^T\beta)$ and $E(\epsilon|X) = E(Y-\mu(X^T\beta)|X) = 0$ by construction. The nuisance tangent space for $F$ in the semiparametric model (\ref{eq:meanmodel})--(\ref{eq:mean}) must therefore be a subspace of the nuisance tangent space for the restricted moment model.

Elementary calculations \citep[see][pp.81--83]{Tsia2006} show that the nuisance tangent space for the restricted moment model is given by
\begin{equation*}
\Lambda = \left\{
\text{all $q \times 1$ functions $a(X,Y)$ such that
$E_{\beta,F}[(Y-\mu(X,\beta))a(X,Y)|X] = 0$}
\right \}
\end{equation*}
and the projection operator $\Pi_{\beta,F}$ onto this nuisance tangent space is given by
\begin{equation*}
\Pi_{\beta,F} s = s - \frac{E_{\beta,F} \left[\left(Y-\mu(X,\beta)\right)s|X\right]}{V(X;\beta,F)} \left(Y-\mu(X,\beta)\right) \ .
\end{equation*}
Applying this operator to the score function (\ref{eq:betascore}), with $\mu(X,\beta) = \mu(X^T\beta)$, gives
\begin{eqnarray*}
\Pi_{\beta,F} S_{\beta,F} &=& X\frac{ \mu'(X^T\beta)}{V(X;\beta,F)} \left(Y-\mu(X^T\beta)\right) \\
& & - X\frac{ \mu'(X^T\beta)}{V(X;\beta,F)} E_{\beta,F} \left[\frac{(Y-\mu(X^T\beta))^2}{V(X;\beta,F)}\Bigg|X \right] (Y-\mu(X^T\beta)) \\
&=& 0 \ \  \mbox{ for all } (\beta, F),
\end{eqnarray*}
because $V(X;\beta,F) = E_{\beta,F}[(Y-\mu(X^T\beta))^2|X]$ by definition. 
Thus, the score function (\ref{eq:betascore}) is orthogonal to the nuisance tangent space in the restricted moment model (\ref{eq:srmm}) and therefore necessarily orthogonal to the nuisance tangent space in the semiparametric model (\ref{eq:meanmodel})--(\ref{eq:mean}) also. 
We summarize as follows:

\begin{Proposition}[Orthogonality]
\label{le:orth}
The mean-model parameter $\beta$ and the error distribution $F$ in any generalized linear model are orthogonal.
\end{Proposition}

Note that the nuisance tangent space for any parametric model (that is, a model in which $F$ is characterized by a finite number of parameters) is necessarily a subspace of the semiparametric nuisance tangent space. We therefore have the following corollary:

\begin{Corollary} [Orthogonality in parametric models]
If the error distribution is characterized by a finite vector of nuisance parameters $\phi$, then the mean-model parameter $\beta$ is orthogonal to $\phi$.
\end{Corollary}

\section{A connection with quasilikelihood models}
\label{sec:4}
A popular extension of GLMs is the class of quasilikelihood (QL) models, also known mean-variance models \citep[e.g.][]{Wed1974}. These models make the assumption that $E(Y|X) = \mu(X,\beta)$ for some mean function $\mu$ and $\mbox{Var}(Y|X) = v(\mu)$ for some positive variance function $v$. Such models can be characterized by their quasi-score functions for $\beta$,
\begin{equation}
\label{eq:quasiscore}
\frac{\partial \mu}{\partial \beta}= \frac{Y-\mu}{v(\mu)} \ .
\end{equation}
In classical QL literature, the functional forms of both $\mu$ and $v$ are usually specified, although there is growing literature on adaptive estimation in which the variance function is left unspecified and estimated nonparametrically from data \citep[e.g.][]{DZ2002}. By comparing score equations (\ref{eq:betascore}) and (\ref{eq:quasiscore}), note that GLMs form a subset of QL models.

For models characterized by (\ref{eq:quasiscore}), \citet{JK2004} showed that the mean-model parameter $\beta$ is orthogonal to the variance function $v$ whenever the latter can be characterized by a finite number of parameters. A general orthogonality result for arbitrary, infinite-dimensional $v$ remains elusive, however, perhaps because of the fact that not all QL score functions (\ref{eq:quasiscore}) correspond to actual probability models. Indeed, such correspondences are atypical. Nevertheless, there is an interesting connection between QL models with unspecified variance functions and GLMs with unspecified error distributions in a certain asymptotic sense made more precise below.

The connection is based on a rather remarkable, but relatively obscure, result from \citet{Hiejima1997}, who showed that GLMs can be considered ``dense" in the class of QL models in the following asymptotic sense: for any mean-variance relationship, there exists an exponential family of distributions (i.e. a GLM with some error distribution $F$) whose score equations for $\beta$ admit roots that are arbitrarily close to the roots of the corresponding QL score equation, as the sample size increases.  Thus, for large enough sample sizes, any adaptive QL model with unspecified variance function $v$ can be approximated arbitrarily well by a GLM with unspecified error distribution $F$, with the latter possessing the orthogonality property \ref{le:orth}. We conjecture that this connection may be the best possible for adaptive QL models, mainly because of the aforementioned fact that QL score functions typically do not correspond to actual probability models.

\section{Practical implications}
\label{sec:5}
The orthogonality property \ref{le:orth} naturally suggests the idea of estimating the error distribution nonparametrically and simultaneously with the mean-model, leading to a kind of ``adaptive GLM". Indeed, if the joint estimation procedure is based on maximum semiparametric likelihood, then the estimator for $\beta$ is guaranteed to be asymptotically efficient and asymptotically independent of the estimated error distribution. In other words, both estimation and inferences on $\beta$ are asymptotically unaffected by having to also estimate the error distribution. In contrast, estimation and inferences in GLMs with misspecified error distributions, or QL models with misspecified variance functions, are generally not efficient. 

The idea of jointly estimating the mean and error distribution in GLMs is considered in more detail in \citet{Huang2013}. In that paper, it is demonstrated that inferences on $\beta$ based on profiling out the error distribution $F$ in the likelihood can be more accurate than inferences based on QL methods. Here, we focus our attention on the accuracy of the  point estimates of $\beta$. The results here complement those found in \citet[][Section 6]{Huang2013}.

\begin{table}
\label{tab:1}
\centering
\caption{\it Relative root mean-square errors of a semiparametric MLE ($\hat \beta_{SP}$) and the usual MLE ($\hat \beta_{MLE}$) in three simulation settings, based on 5000 replications each.
}
\small{
\begin{tabular}{lllccccccc}
\hline
 & & & \uline{\hspace{3mm}Esti\hspace{-2.5mm}} & \uline{\hspace{-2.5mm}mator\hspace{3mm}} \\
Data &n	& Parameter	& $\hat \beta_{SP}$ & $\hat \beta_{MLE}$ \\
\hline
Exponential & 33 & Intercept 	& 0.199 & 0.194 \\
& & Group effect 			 	  & 0.361  & 0.354 \\
& & Common slope  		 & 0.464 & 0.455 \\
& 66 & Intercept 			 &  0.124 & 0.122  \\
& & Group effect 			 & 0.247 & 0.243 \\
& & Common slope  	 	& 0.300 & 0.297 \\
Poisson & 44 & Intercept & 0.275 & 0.271 \\
 & & Coefficient of $X_1$ & 0.610 & 0.594 \\
 & & Coefficient of $X_2$ & 0.205 & 0.201\\
 & & Coefficient of $X_3$ & 0.546& 0.533\\
\hline
\end{tabular}
}
\end{table}

In Table 1, we compare the relative root mean-square error of a semiparametric MLE of $\beta$ (with $F$ unknown) to that of the usual MLE (with $F$ set to the true distribution) from three sets of simulations. Recall that the relative root mean-square error of an estimator $\hat \beta$ is defined as the root mean-square error of $\hat \beta$ divided by the absolute value $|\beta|$ of the parameter. The simulation settings are based on a leukemia survival dataset from \citet{DH1997} and a mine injury dataset from \citet{MMVR2010}, and are described in more detail in Sections 6.1 and 6.2 of \citet{Huang2013}. The particular semiparametric estimation approach we use for estimating $\beta$ 
is based on empirical likelihood and is described in more detail in Section 4 of \citet{Huang2013}. 

We see from Table 1 that the relative root mean-square errors of the two estimators are essentially the same, even for moderately small sample sizes. This supports the claim that maximum likelihood estimation of $\beta$ is asymptotically efficient regardless of whether $F$ is known or completely unknown and estimated nonparametrically from data.

\section{Conclusion}
In this note, we have shown that orthogonality between the mean-model parameter and the error distribution holds for any GLM, parametric or nonparametric. This confirms, in greatest generality, what is well-known for the special cases of normal, gamma and multinomial regression. The result also has implications for applied statistical work, with our numerical results suggesting that little is lost by treating the error distribution nonparametrically, even in moderately sized problems. (It can also be said that little is gained by knowing the error distribution completely!) Nonparametric estimation of the error distribution can therefore safeguard against biases due to parametric model misspecification, without sacrificing much in terms of efficiency.

\section*{Acknowledgments}
The authors thank the Associate Editor and an anonymous referee for suggestions that improved the paper. Paul. J. Rathouz was funded by NIH grant R01 HL094786. 

\appendix


\small
\begin{thebibliography}{99}\setlength{\itemsep}{0.0mm}

\bibitem[Cox \& Reid(1987)]{CR1987}
Cox, D. R. and Reid, N. (1987). Parameter orthogonality and approximate conditional inference, {\it J. Roy. Stat. Soc. Ser. B}, 49, 1--39.

\bibitem[Davison \& Hinkley(1997)]{DH1997}
Davison, A. C. and Hinkley, D. V. (1997). {\it Bootstrap methods and their applications}, Cambridge University Press, Cambridge, UK.

\bibitem[Dewanji \& Zhao(2002)]{DZ2002}
Dewanji, A. and Zhao, L. P. (2002). An optimal estimating equation with unspecified variances {\it Sankhy\={a}}, 64, 95--108.

\bibitem[Hiejima(1997)]{Hiejima1997}
Hiejima, Y. (1997). Interpretation of the quasi-likelihood via the tilted exponential family, {\it J. Japan Statist. Soc}, 27, 157--164.


\bibitem[Huang(2013)]{Huang2013}
Huang, A. (2013). Joint estimation of the mean and error distribution in generalized linear models, {\it J. Amer. Stat. Assoc.}, 
(to appear).

\bibitem[J{\o}rgensen(1987)]{Jorg1987}
J{\o}rgensen, B. (1987). Exponential dispersion models, {\it J. Roy. Stat. Soc. Ser. B}, 49, 127--162, with discussion and a reply by the author.

\bibitem[J{\o}rgensen \& Knudsen(2004)]{JK2004}
J{\o}rgensen, B. and Knudsen, S. J. (2004). Parameter orthogonality and adjustment for estimating functions{\it. Scand. J. Stat. }, 31, 93--114.

\bibitem[McCullagh \& Nelder(1989)]{MN1989}
McCullagh, P. and Nelder, J. A. (1989). {\it Generalized linear models}, Chapman \& Hall, London, UK.

\bibitem[Myers et al.(2010)]{MMVR2010}
Myers, R. H., Montgomery, D. C., Vining, G. G., and Robinson, T. J. (2010). {\it Generalized linear models},
John Wiley \& Sons Inc., Hoboken, USA.

\bibitem[Rathouz \& Gao(2009)]{RG2009}
Rathouz, P. J. and Gao, L. P. (2009). Generalized linear models with unspecified reference distribution, {\it Biostatistics}, 10, 205--218.


 \bibitem[Tsiatis(2006)]{Tsia2006}
Tsiatis, A. A. (2006). {\it Semiparametric theory and missing data}, Springer, New York, USA.

\bibitem[Wang \& Famoye(1997)]{WF1997}
Wang, W. and Famoye, F. (1997). Modeling household fertility decisions with generalized Poisson regression, {\it J. Population. Econ.}, 10, 273--283.

\bibitem[Wedderburn(1974)]{Wed1974}
Wedderburn, R. W. M. (1974). Quasi-likelihood functions, generalized linear models, and the Gauss-Newton method, {\it Biometrika}, 61, 439--447.

\end{thebibliography}
\end{document}